# Title: A giant exoplanet orbiting a very-low-mass star challenges planet formation models

**Authors:** J. C. Morales[1,2*], A. J. Mustill[3], I. Ribas[1,2], M. B. Davies[3], A. Reiners[4], F. F. Bauer[5], D. Kossakowski[6], E. Herrero[1,2], E. Rodríguez[5], M. J. López-González[5], C. Rodríguez-López[5], V. J. S. Béjar[7,8], L. González-Cuesta[7,8], R. Luque[7,8], E. Pallé[7,8], M. Perger[1,2], D. Baroch[1,2], A. Johansen[3], H. Klahr[6], C. Mordasini[9], G. Anglada-Escudé[10,5], J. A. Caballero[11], M. Cortés-Contreras[11], S. Dreizler[4], M. Lafarga[1,2], E. Nagel[12], V. M. Passegger[13], S. Reffert[13], A. Rosich[1,2], A. Schweitzer[12], L. Tal-Or[4,4], T. Trifonov[6], M. Zechmeister[4], A. Quirrenbach[13], P. J. Amado[5], E.W. Guenther[15], H.-J. Hagen[12], T. Henning[6], S. V. Jeffers[4], A. Kaminski[13], M. Kürster[6], D. Montes[16], W. Seifert[13], F. J. Abellán[17,16], M. Abril[5], J. Aceituno[18,5], F. J. Aceituno[5], F. J. Alonso-Floriano[19,16], M. Ammler-von Eiff[20,15], R. Antona[5], B. Arroyo-Torres[18], M. Azzaro[18], D. Barrado[11], S. Becerril-Jarque[5], D. Benítez[18], Z. M. Berdiñas[21,5], G. Bergond[18], M. Brinkmöller[13], C. del Burgo[22], R. Burn[9], R. Calvo-Ortega[5], J. Cano[16], M. C. Cárdenas[6], C. Cardona Guillén[7,8], J. Carro[16], E. Casal[5], V. Casanova[5], N. Casasayas-Barris[7,8], P. Chaturvedi[15], C. Cifuentes[11,16], A. Claret[5], J. Colomé[1,2], S. Czesla[12], E. Díez-Alonso[16,23], R. Dorda[16,7,8], A. Emsenhuber[24], M. Fernández[5], A. Fernández-Martín[18], I. M. Ferro[5], B. Fuhrmeister[12], D. Galadí-Enríquez[18], I. Gallardo Cava[25,16], M. L. García Vargas[26], A. Garcia-Piquer[1,2], L. Gesa[1,2], E. González-Álvarez[27], J. I. González Hernández[7,8], R. González-Peinado[16], J. Guàrdia[1,2], A. Guijarro[18], E. de Guindos[18], A. P. Hatzes[15], P. H. Hauschildt[12], R. P. Hedrosa[18], I. Hermelo[18], R. Hernández Arabi[18], F. Hernández Otero[18], D. Hintz[12], G. Holgado[7,8,16], A. Huber[6], P. Huke[4], E. N. Johnson[4], E. de Juan[18], M. Kehr[15], J. Kemmer[13], M. Kim[28,13], J. Klüter[29,13], A. Klutsch[30,16], F. Labarga[16], N. Labiche[13], S. Lalitha[4], M. Lampón[5], L. M. Lara[5], R. Launhardt[6], F. J. Lázaro[16], J.-L. Lizon[31], M. Llamas[16], N. Lodieu[7,8], M. López del Fresno[11], J. F. López Salas[1,2], J. López-Santiago[32,33,16], H. Magán Madinabeitia[5,18], U. Mall[6], L. Mancini[34,6,35,36], H. Mandel[13], E. Marfil[16], J. A. Marín Molina[18], E. L. Martín[27], P. Martín-Fernández[18], S. Martín-Ruiz[5], H. Martínez-Rodríguez[37,16], C. J. Marvin[4], E. Mirabet[5,12], A. Moya[38,39,11], V. Naranjo[6], R. P. Nelson[10], L. Nortmann[7,8], G. Nowak[7,8], A. Ofir[40], J. Pascual[5], A. Pavlov[6], S. Pedraz[18], D. Pérez Medialdea[5], A. Pérez-Calpena[26], M. A. C. Perryman[41], O. Rabaza[42,5], A. Ramón Ballesta[5], R. Rebolo[7,8], P. Redondo[7], H.-W. Rix[6], F. Rodler[43,1,2], A. Rodríguez Trinidad[5], S. Sabotta[15], S. Sadegi[13,6], M. Salz[12], E. Sánchez-Blanco[44], M. A. Sánchez Carrasco[5], A. Sánchez-López[5], J. Sanz-Forcada[11], P. Sarkis[6], L. F. Sarmiento[4], S. Schäfer[4], M. Schlecker[6], P. Schöfer[4], E. Solano[11], A. Sota[5], O. Stahl[13], S. Stock[13], T. Stuber[13], J. Stürmer[45,13], J. C. Suárez[46,5], H. M. Tabernero[27], S. M. Tulloch[47], G. Veredas[13], J. I. Vico-Linares[18], F. Vilardell[1,2], K. Wagner[13], J. Winkler[15], V. Wolthoff[13], F. Yan[4], M. R. Zapatero Osorio[27]

**Affiliations:**
[1] Institut de Ciències de l'Espai (ICE,CSIC), Campus UAB,C/ de Can Magrans s/n, E-08193 Bellaterra, Spain
[2] Institut d'Estudis Espacials de Catalunya (IEEC), C/ Gran Capità 2-4, E-08034 Barcelona, Spain
[3] Lund Observatory, Department of Astronomy & Theoretical Physics, Lund University, Box 43, SE-221 00 Lund, Sweden
[4] Institut für Astrophysik, Georg-August-Universität, Friedrich- Hund-Platz 1, D-37077 Göttingen, Germany
[5] Instituto de Astrofísica de Andalucía (IAA-CSIC), Glorieta de la Astronomía s/n, E-18008

Granada, Spain

[6] Max-Planck-Institut für Astronomie, Königstuhl 17, D-69117 Heidelberg, Germany

[7] Instituto de Astrofísica de Canarias, Vía Láctea s/n, E-38205 La Laguna, Tenerife, Spain

[8] Departamento de Astrofísica, Universidad de La Laguna, E-38206 La Laguna, Tenerife, Spain

[9] Physikalisches Institut, Universität Bern, Gesellschaftsstrasse 6, CH-3012 Bern, Switzerland

[10] School of Physics and Astronomy, Queen Mary University of London, 327 Mile End Rd, E1 4NS London, United Kingdom

[11] Centro de Astrobiología (CSIC-INTA), Campus ESAC (ESA), Camino Bajo del Castillo s/n, E-28692 Villanueva de la Cañada, Spain

[12] Hamburger Sternwarte, Universität Hamburg, Gojenbergsweg 112, D-21029 Hamburg, Germany

[13] Landessternwarte, Zentrum für Astronomie der Universität Heidelberg, Königstuhl 12, D-69117 Heidelberg, Germany

[14] Department of Geophysics, Raymond and Beverly Sackler Faculty of Exact Sciences, Tel Aviv University, Tel-Aviv 6997801, Israel

[15] Thüringer Landessternwarte Tautenburg, Sternwarte 5, D-07778 Tautenburg, Germany

[16] Departamento de Física de la Tierra y Astrofísica & IPARCOS-UCM (Instituto de Física de Partículas y del Cosmos de la UCM), Facultad de Ciencias Físicas, Universidad Complutense de Madrid, E-28040 Madrid, Spain

[17] Departamento de Astronomía y Astrofísica, Universidad de Valencia, C/Dr. Moliner 50, E-46100 Burjassot, Spain

[18] Centro Astronómico Hispano-Alemán (CSIC-MPG), Observatorio Astronómico de Calar Alto, Sierra de los Filabres, E-04550, Gérgal, Almería, Spain

[19] Leiden Observatory, Leiden University, Postbus 9513, 2300 RA, Leiden, The Netherlands

[20] Max Planck Institute for Solar System Research, Justus-von-Liebig-Weg 3, D-37077 Göttingen, Germany

[21] Departamento de Astronomía, Universidad de Chile, Camino El Observatorio, 1515 Las Condes, Santiago, Chile

[22] Instituto Nacional de Astrofísica, Óptica y Electrónica, Luis Enrique Erro 1, Sta. Ma. Tonantzintla, Puebla, Mexico

[23] Departamento de Explotación y Prospeción de Minas, Escuela de Minas, Energía y Materiales, Universidad de Oviedo, E-33003 Oviedo, Asturias, Spain

[24] Lunar & Planetary Laboratory, University of Arizona, 1629 E. University Blvd., Tucson AZ 85721, USA

[25] Observatorio Astronómico Nacional (OAN-IGN), Apartado 112, E-28803 Alcalá de Henares, Spain

[26] FRACTAL S.L.N.E. C/ Tulipán 2, portal 13, 1A. E-28231 Las Rozas de Madrid (Spain)

[27] Centro de Astrobiología (CSIC-INTA), Carretera de Ajalvir km 4, E-28850 Torrejón de Ardoz, Madrid, Spain

[28] Institut für Theoretische Physik und Astrophysik, Leibnizstraße 15, D-24118 Kiel, Germany

[29] Zentrum für Astronomie der Universität Heidelberg, Astronomisches Rechen-Institut, Mönchhofstr. 12-14, D-69120 Heidelberg Germany

[30] Institut für Astronomie und Astrophysik, Eberhard Karls Universität, Sand 1, D-72076 Tübingen, Germany

[31] European Organisation for Astronomical Research in the Southern Hemisphere, Karl-Schwarzschild-Str. 2, D-85748 Garching bei München, Germany

[32] Department of Signal Theory and Communications, Universidad Carlos III de Madrid, Av. de la Universidad 30, E-28911, Leganés, Madrid, Spain

[33] Gregorio Marañón Health Research Institute, Doctor Esquerdo 46, E28007, Madrid, Spain

[34] Department of Physics, University of Rome Tor Vergata, Via della Ricerca Scientifica 1, I-00133, Roma, Italy

[35] INAF - Osservatorio Astrofisico di Torino, via Osservatorio 20, I-10025, Pino Torinese, Italy

[36] International Institute for Advanced Scientific Studies (IIASS), Via G. Pellegrino 19, I-84019, Vietri sul Mare (SA), Italy

[37] Department of Physics and Astronomy and Pittsburgh Particle Physics, Astrophysics and Cosmology Center (PITT PACC), University of Pittsburgh, 3941 O'Hara Street, Pittsburgh, PA 15260, USA

[38] School of Physics and Astronomy, University of Birmingham, Edgbaston, Birmingham, B15 2TT, UK

[39] Stellar Astrophysics Centre, Department of Physics and Astronomy, Aarhus University, Ny Munkegade 120, DK-8000 Aarhus C, Denmark

[40] Department of Earth and Planetary Sciences, Weizmann Institute of Science, Rehovot, 76100, Israel

[41] School of Physics, University College Dublin, Belfield Downs, Dublin, D14 YH57, Ireland

[42] Dpto. Ingeniería Civil, Universidad de Granada, Campus de Fuentenueva s/n, E-18071, Granada, Spain

[43] European Southern Observatory (ESO), Alonso de Córdova 3107, Vitacura, Casilla 19001, Santiago de Chile

[44] Diseño de Sistema Ópticos, María Moliner 9B, E-41008 Sevilla, Spain

[45] The Department of Astronomy and Astrophysics, University of Chicago, 5640 S. Ellis Ave, Chicago, IL, 60637

[46] Dpto. Física Teórica y del Cosmos, Universidad de Granada, Campus de Fuentenueva s/n, E-18071, Granada, Spain

[47] University of Sheffield, Dept. of Physics and Astronomy, Sheffield S3 7RH, United Kingdom

*Correspondence to: morales@ice.cat

## Abstract

Statistical analyses from exoplanet surveys around low-mass stars indicate that super-Earth and Neptune-mass planets are more frequent than gas giants around such stars, in agreement with core accretion theory of planet formation. Using precise radial velocities derived from visual and near-infrared spectra, we report the discovery of a giant planet with a minimum mass of 0.46 Jupiter masses in an eccentric 204-day orbit around the very low-mass star GJ 3512. Dynamical models show that the high eccentricity of the orbit is most likely explained from planet-planet interactions. The reported planetary system challenges current formation theories and puts stringent constraints on the accretion and migration rates of planet formation and evolution models, indicating that disc instability may be more efficient in forming planets than previously thought.

**One Sentence Summary**

A Jupiter-mass planet found orbiting the low-mass star GJ 3512 favours a scenario of formation by disk instability.

**Main Text**

Almost 4000 exoplanets have been discovered to date, but only about 10% are orbiting low-mass M dwarf stars, in spite of this stellar type being the most numerous in the Galaxy. This observational bias is largely a consequence of the intrinsic faintness of M dwarfs at visual wavelengths, where most exoplanet searches have been conducted. Statistical studies, based on radial-velocity and transit surveys (*1*, *2*), yield estimates between 1 and 2.5 planets per M dwarf, most of them in the Earth- and Neptune-mass regime (*3*). Only a few Jupiter-mass planets have been found to orbit late-type stars (*4*, *5*). This is consistent with the core accretion theory of planet formation (*6*, *7*), which predicts a low abundance of gas giants orbiting such stars. Alternative theories, such as disc instability, may explain the formation of gas giant planets in high-mass protoplanetary discs (*8*, *9*). Microlensing survey results (*10*, *11*) indicate that gas giant planets may be more abundant at larger distances from their host stars, where transit and radial velocity surveys are less sensitive. This would be in agreement with exoplanet formation scenarios suggesting that the occurrence of gas giant planets may increase beyond the snow line (the distance from the star beyond which volatile compounds could condense) of protoplanetary discs, but it is not yet clear if discs around late-type stars have sufficient material and survive long enough to form such massive planets (*6*, *12*). The CARMENES exoplanet survey (*13*) aims to answer this open question by searching for exoplanets around M-dwarf stars using a dual channel high-resolution spectrograph operating in the visible and near-infrared wavelength ranges. Entering its fourth year of survey, CARMENES is now able to detect planets that reside beyond the snow line of their host stars, which is closer for late-type M dwarfs. Therefore, the radial velocity signals of planetary companions are larger and the orbital periods shorter (from a few hundred days).

GJ 3512 (LP 90-18) is a high-proper motion late-type star included in the CARMENES survey whose basic properties are summarised in Table 1. It is classified as an M5.5 main-sequence star, with a mass of about 0.12 $M_\odot$, and the Gaia mission provides a trigonometric distance of 9.489 pc (*14*). Stellar properties, i.e. mass, radius, effective temperature, luminosity, and metallicity, were computed from the spectroscopic observations using the same procedures developed for other CARMENES survey stars (*15*). The first routine observations of this target showed a clear trend in radial velocities, which prompted an increase in the observational cadence. Over the past two years we have secured full coverage of a large-amplitude radial velocity periodic modulation. Fig. 1 shows the radial velocity curve derived from the visual and near-infrared channels of the CARMENES instrument. A Keplerian model provides parameters in agreement for both wavelength channels, confirming that the variability is consistent with a planet interpretation. A simultaneous fit yields a planet with a minimum mass of $0.463^{+0.022}_{-0.023}$ Jupiter masses in an eccentric (*e*=0.4356±0.0042) orbit with a period of 203.59 days. The residuals of the best fit were inspected for additional variability, showing a significant long-term trend in the data, also visible in both wavelength channels. Adding a second-order polynomial to the Keplerian motion significantly improves the radial velocity fit with respect to a linear model. This indicates that the residual radial velocities could be the reflex radial velocity of a second object with a period longer than ~1400 days. Therefore, we finally adopted a two-planet Keplerian model as the best fit to the data, assuming a circular orbit for the poorly-constrained, long-period signal (see Fig. 1 and *16*). No further significant signals are identified in the dataset. Table 1 lists the parameters of the final fit, along with the orbital and planetary properties derived for GJ 3512 b and the constraints for the candidate GJ 3512 c. The uncertainties are computed using standard Markov Chain Monte Carlo (MCMC) procedures (*17*) and considering 68% credibility intervals.

GJ 3512 is a moderately magnetically-active star showing emission in the Hα line (*18*). Photometric monitoring of the target conducted from the Sierra Nevada, Montsec, and Las Cumbres observatories yielded light curves showing variability with a period of around 87 days and with a peak-to-peak modulation of about 3%. Activity indices such as the differential line width computed from the spectra (*19*) also show significant signals around the same period. We attribute this variability to modulation caused by the rotation of the star, with a period that we estimate to be 87±5 days (*16*), in agreement with ref. (*20*), who report a value of ~87 days. The relatively long rotation period, and the measured space motions indicate that GJ 3512 is a rather old star with a most likely age in the interval 3-8 Ga (*16*). This is also consistent with its approximately solar metallicity. An explanation of the 204-day radial-velocity variability as arising from magnetic activity can be confidently ruled out because this period is not present in the activity indicators, because the radial velocity amplitude is larger than expected for a moderately-active slow-rotating star such as GJ 3512, and also because no dependence of the radial velocity amplitude with wavelength is seen from the visual to the near-infrared.

Although the orbital inclination is unknown, only values below 2 deg, i.e., an orbit nearly face-on, would result in an absolute mass above the brown dwarf limit (~13 Jupiter masses) for GJ 3512 b. This means that the probability of this object being a planetary-mass body is very high (>99.9%). Fig. 2 shows the location of GJ 3512 b in the minimum-mass versus host star mass plot compared to known planetary systems. It lies in a region of the parameter space corresponding to low-mass stars with massive planetary companions, which has not been explored in detail because of the faintness of the targets and the very low transit probability in wide orbits (<0.5% for GJ 3512 b). GJ 3512 is the lowest mass star with a giant planet detected so far by radial velocities, with a minimum mass ratio $q \sim 0.0034$. This system lies in the parameter space region where only the microlensing technique has reported planet discoveries so far (*10*, *21*), but the transient nature of the detection and the very large uncertainties of the host star masses prevent a deeper individual analysis. The orbital and physical properties of the exoplanet GJ 3512 b, as well as of the host star, are determined much more accurately. It is interesting to note that GJ 3512 is a twin of the nearby planet-hosting star Proxima Centauri (*22*), with very similar stellar parameters but with completely different planetary architectures. Preliminary statistical estimates from the CARMENES survey and the literature indicate that the occurrence rate of giant planets in orbits up to a few astronomical units around stars with masses below 0.3 $M_{\odot}$ is about 3%, this value being compatible with microlensing surveys (*23*).

The high orbital eccentricity of GJ 3512 b is not expected for a system with only one planet, for which the interaction with the stellar disc during migration should lead to a circular or low-eccentricity orbit. However, planet–planet scattering has been shown to be a possible way of explaining the often-large eccentricities of giant planets (*7*). Given the likely existence of a second wide-orbit planet, inferred from the trend in the radial velocity residuals after subtracting planet b, we find that a plausible route to the present orbital architecture is that the system formed initially with three planets, of which one (with a mass similar to or lower than planet b) was ejected, leaving GJ 3512 b on an eccentric orbit and a large gap between the two surviving planets (*16*).

Moreover, GJ 3512 b has a very high mass for such a small host star, which, combined with the probably high planetary multiplicity at birth, poses a significant challenge for planet formation theories. We explored planet formation scenarios around GJ 3512 with the latest pebble accretion models (*24*), without success (*16*). Formation of a gas giant in this way requires building up a large planetary core of at least 5 Earth masses. We show here that this is not possible around such a low-mass M dwarf because the migration rate of planets around low-mass stars is high. Therefore, planetary seeds move rapidly to the inner edge of the disc. Assuming high disc masses is not favoured by the observations (*25*) but also does not resolve the issue because it leads simultaneously to higher accretion and migration rates (*16*). Furthermore, allowing for the possibility of a longer disc lifetime around low mass stars (*26*) does not help.

We note that the total mass of the planets ($\gtrsim$0.5 $M_J$, >0.005 $M_\star$, including at least GJ 3512 b and c) is significant compared to the range of disc masses around low-mass pre-main sequence Class II M dwarfs (~0.1–10 $M_J$) (*27*), implying an extremely high planet formation efficiency. We therefore turned to the competing model of planet formation by gravitational instability in the gas disc at very young ages, when the disc is still massive relative to the star ($M_{\mathrm{disc}}/M_\star \gtrsim 0.1$) (*28, 29*). For a range of disc viscosities $\alpha$ and surface densities $\Sigma$, the disc is gravitationally unstable at radii < 100 au (Fig. 3 and *16*). Furthermore, the estimated masses of the fragments formed (*30*) are less than that of Jupiter, which is in line with the mass of GJ 3512 b. Except for unrealistically low values of $\alpha$, the disc fragments at radii $\gtrsim$10 au, and so the planets must have migrated a significant distance from their formation location to their present position. This is not a problem given the large mass of the disc with respect to the planet, and it is indeed often seen in numerical simulations of disc fragmentation (*30, 31*). In the region of realistic viscosity $\alpha$ > 0.01, the disc typically fragments at radii of a few tens of au, and the total disc mass within this radius is ~30 $M_J$. Discs cannot extend too far beyond this fragmentation radius, since the total disc mass becomes extremely large (up to 1 $M_\odot$ within 100 au). Thus, the planetary system around GJ 3512 favours the gravitational instability scenario as the formation channel for giant planets around very low-mass stars.

Given the mass ratio between GJ 3512 b and its host star, and the large orbital semi-major axis, the Gaia mission will provide further information for this system. Assuming the orbital properties reported in Table 1, the reflex motion of the host star is expected to be around 120 μas, which is about two times larger than the expected accuracy of Gaia astrometry (*32*) for $G \sim$ 13 mag. Actually, the Gaia second data release reports an excess noise in the astrometric five-parameter fit of 632 μas, which is highly significant and may be partially attributed to the inner planetary companion. Assuming that all of the astrometric excess noise is due to the inner planet, this provides a lower limit for the inclination and an upper limit for the planet mass of about 8 deg and ~3.5 Jupiter masses, respectively. Therefore, one can safely exclude any face-on configurations where the companion would be considerably more massive than a few Jupiter masses or in the brown dwarf regime. Eventually, Gaia astrometry should provide the inclination and mass of the inner planet as well as possibly the orbital parameters of the outer companion, and their mutual inclination, further constraining planet-planet scattering models. Furthermore, the star-planet contrast is estimated to be $\sim 10^{-7}$ only (depending on the albedo, inclination, and planet size), and is thus potentially imageable with future instruments. This makes GJ 3512 a very promising system because it may be fully characterized and thus continue to place stringent constraints on accretion and migration processes, as well as on the efficiency of planet formation in protoplanetary discs, and the disc-to-star mass ratios.

**Acknowledgments:** These results were based on observations made with the CARMENES instrument at the 3.5-m telescope of the Centro Astronómico Hispano-Alemán de Calar Alto (CAHA, Almería, Spain), funded by the German Max-Planck-Gesellschaft (MPG), the Spanish Consejo Superior de Investigaciones Científicas (CSIC), the European Union, and the CARMENES Consortium members, and stored at the CARMENES data archive at CAB (INTA-CSIC), the 150-cm and the 90-cm telescope at the Sierra Nevada Observatory (Granada, Spain), operated by the Instituto de Astrofísica de Andalucía (IAA-CSIC), and the 80-cm Joan Oró Telescope (TJO) of the Montsec Astronomical Observatory (OAdM), owned by the Generalitat de Catalunya and operated by the Institute for Space Studies of Catalonia (IEEC). **Funding:** This research was supported by the following programmes, grants and fellowships: Spanish Ministry for Science, Innovation and Universities (MCIU) ESP2014-54062-R, ESP2014-54362P, AYA2015-69350-C3-2-P, BES-2015-074542, AYA2016-79425-C3-1/2/3-P, ESP2016-76076-R, ESP2016-80435-C2-1-R, ESP2016-80435-C2-2-R, ESP2017-87143-R, ESP2017-87676-C05-02-R, ESP2017-87676-2-2, and RYC-2012-09913 ('Ramón y Cajal' programme); Israel Science Foundation grant No. 848/16; CONICYT-FONDECYT/Chile Postdoctorado 3180405; Deutsches Zentrum für Luft- un Raumfahrt (DLR) 50OW0204 and 50OO1501; Italian Minister of Instruction, University and Research (MIUR), FFABR 2017; University of Rome Tor Vergata, "Mission: Sustainability 2016" fund; European Research Council under the European Union Horizon 2020 reasearch and innovation programe No. 694513; Mexican national council for science and technology CONACYT, CVU No. 448248; the "Center of Excellence Severo Ochoa" award for the Instituto de Astrofísica de Andalucía (SEV-2017-0709); Generalitat de Catalunya/CERCA programme; Fondo Europeo de Desarrollo Regional (FEDER); German Science Foundation (DFG) Research Unit FOR2544 "Blue Planets around Red Stars" and Priority Programs SPP 1833 "Building a Habitable Earth", SPP 1992 "Exploring the Diversity of Extrasolar Planets"; National Science Foundation under Grants No. NSF PHY17-48958 and PHY-1607761; Swiss National Science Foundation under grant BSGI0_155816 "PlanetsInTime" and within the framework of the NCCR PlanetS; Queen Mary University of London Scholarship; Spanish MCIU FPI-SO predoctoral contract BES-2017-082610; and the Knut and Alice Wallenberg Foundation.

**Author contributions**

J.C.M. organized and participated in the analysis of the spectroscopic and photometric data and their interpretation, and wrote most of the manuscript. A.J.M, M.B.D., A.J., H.K., and C.M. discussed the formation mechanism of this exoplanet system and performed the numerical simulations. I.R. co-led the analysis and contributed to the preparation of the manuscript. A.Re. co-led the CARMENES contribution and discussed the characterization of the system. F.F.B. and D.K. analysed the different orders of CARMENES near-infrared spectra to increase the precision of radial-velocity data. E.H. obtained, reduced and analysed the TJO photometry. E.R., M.J.L-G., and C.R-L. coordinated the photometric follow-up of the targets at SNO observatory and analysed the SNO photometry. F.J.Ac., V.C., and A.So. collected the SNO photometry. V.J.S.B., L.G-C., R.Lu., and E.P. coordinated the photometric follow-up of the targets and obtained and reduced data from Las Cumbres Observatory. M.P., G.A.-E., T.T., and A.Ro. contributed to the analysis of the radial velocities. D.Baro. performed a two dimensional cross-correlation function to discard stellar companions in the system and revised known exoplanet properties. M.Laf. calculated the cross-correlation function parameters of CARMENES spectra. J.A.C., M.C-C., V.M.P. and A.Sc. derived the basic stellar parameters. L.T.-O. corrected and calibrated the performance of CARMENES data. M.Z. reduced the CARMENES data. A.Q. and P.J.A. are principal investigators of CARMENES. A.E., M.S., and R.M. contributed to the discussion and simulation of the formation scenarios. E.S. and M.L.F. are responsible for the CARMENES GTO archive. All authors but A.J.M., M.B.D., A.J., C.M., F.J.Ac., V.C., R.B., A.E., R.B., and A.So. have contributed to the building and operation of the CARMENES instrument.

| Star parameter | Value |
|---|---|
| Spectral type | M5.5 V |
| Mass ($M_\odot$) | 0.123±0.009 |
| Radius ($R_\odot$) | 0.139±0.005 |
| Luminosity ($L_\odot$) | 0.00157±0.00002 |
| Effective temperature (K) | 3081±51 |
| Distance (pc) | 9.489±0.008 |
| Rotation period (d) | 87±5 |
| Space velocities (km s$^{-1}$) | U=−28.80±0.37, V=−51.55±0.17, W=1.51±0.30 |
| Metallicity | −0.07±0.16 |

| Planet parameter | Value | |
|---|---|---|
| | GJ 3512 b | GJ 3512 c |
| Orbital period (d) | $203.59^{+0.14}_{-0.14}$ | >1390 |
| Radial velocity semi-amplitude (m s$^{-1}$) | $71.84^{+0.34}_{-0.36}$ | >12 |
| Eccentricity | $0.4356^{+0.0042}_{-0.0042}$ | … |
| Argument of periastron (deg) | $125.49^{+0.71}_{-0.73}$ | … |
| Time of periastron - BJD2450000.0 (d) | $7745.65^{+0.50}_{-0.50}$ | … |
| Minimum mass ($M \sin i$; $M_J$) | $0.463^{+0.022}_{-0.023}$ | >0.17 |
| Orbital semi-major axis (au) | $0.3380^{+0.0080}_{-0.0084}$ | >1.2 |
| Minimum astrometric semi-amplitude ($\alpha \sin i$; mas) | $0.12753^{+0.00062}_{-0.00065}$ | >0.15 |
| Maximum planet angular separation (mas), $i = 90$ deg | $42.2^{+1.0}_{-1.1}$ | >120 |
| $i = 8$ deg | $51.1^{+1.2}_{-1.3}$ | >120 |

**Table 1. Information of GJ 3512 and its planet candidates.** We derived fundamental stellar parameters of GJ 3512 as in ref. *15*. Orbital and planetary parameters, and their uncertainties, are determined by calculating the mean values and 68% credibility intervals of the distribution resulting from the MCMC run. Only lower limits can be obtained for GJ 3512 c. Complementary fitting algorithms, e.g., using a Gaussian Processes framework, were also employed in the modelling of the data and yielded compatible results within the uncertainties.

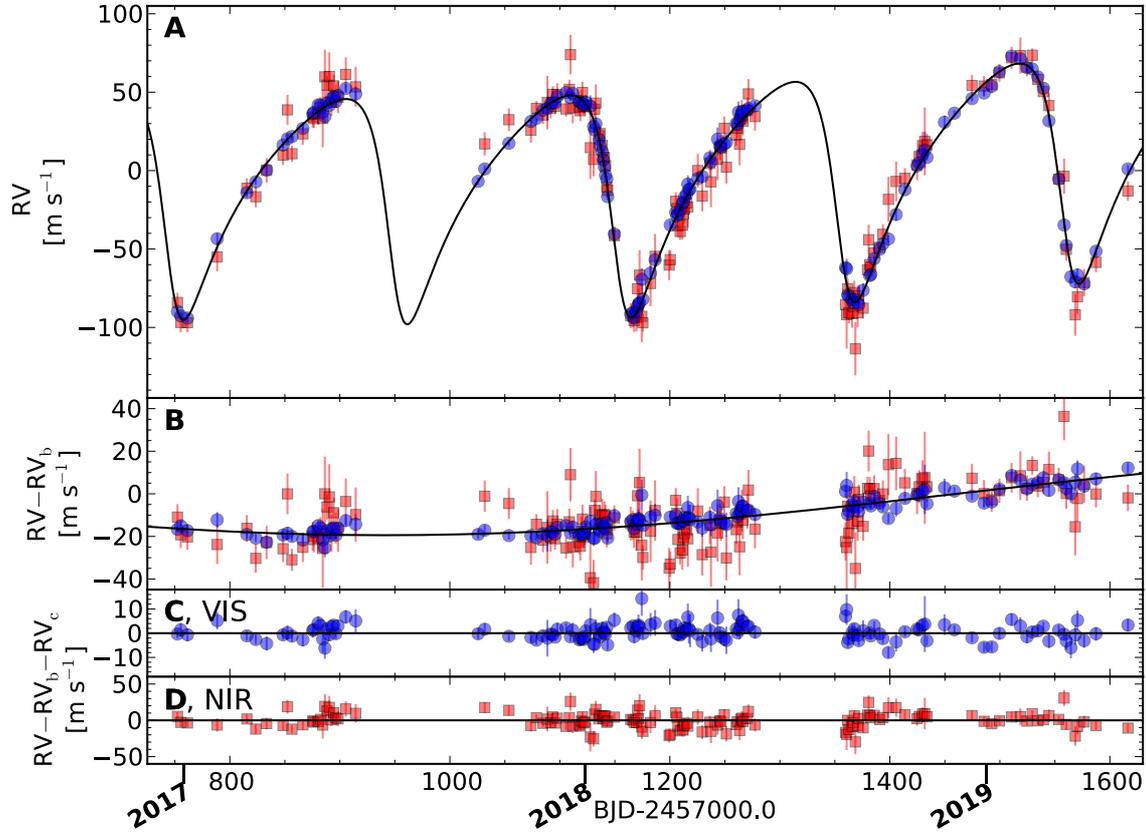

**Fig. 1. Time-series of radial velocity data and residuals.** (A) the radial velocity time-series obtained with the CARMENES visual (blue circles) and near-infrared (red squares) channels, and the best-fitting Keplerian orbital model (black solid line). (B) the same data and fits after removing the signal of the inner planet GJ 3512 b (RV$_b$), showing the radial velocity of the long-period candidate GJ 3512 c (RV$_c$). (C) and (D) the residuals between the best fitting two-planet model and data for the two CARMENES channels. This model includes two Keplerian orbits, with the longer-period orbit assumed to be circular with $P$=2100 days, which yields the best likelihood value. Black horizontal lines are guidelines, not fits to the data. Each panel has a different vertical scale and horizontal axis is the time of the observations in barycentric Julian day (BJD). Calendar years are indicated for reference.

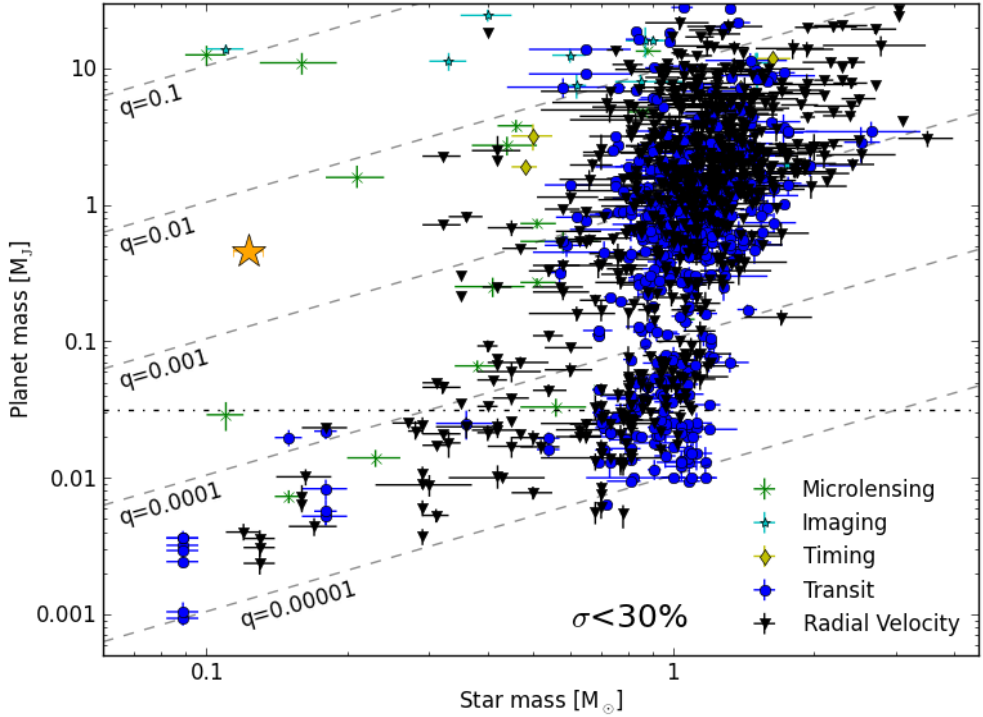

**Fig. 2. GJ 3512 b minimum planet mass and host stellar mass compared to known planetary systems.** Data for known exoplanets come from the NASA exoplanet archive. Only systems with star and planet mass uncertainties below 30% are displayed (see Fig. S4 for a full comparison with all planetary systems). Different exoplanet detection techniques are shown as labelled and GJ 3512 b is depicted with an orange star symbol. The planet minimum mass is plotted in the case of planets detected by radial velocities and timing. The dashed lines indicate different host star-to-planet mass ratios (*q*) as labelled, and the horizontal dot-dashed line corresponds to 10 M⊕.

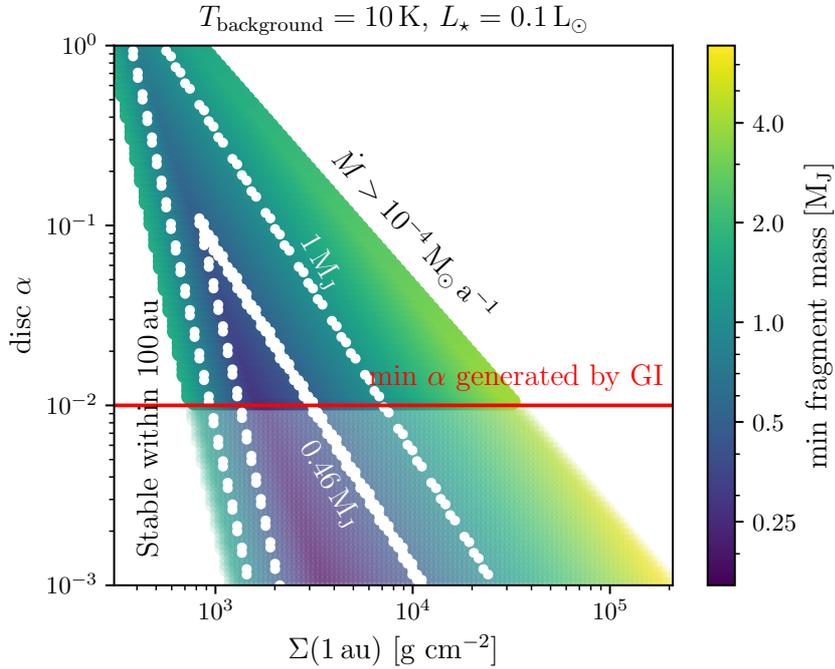

**Fig. 3. Planet formation around GJ 3512 through gravitational instability.** Gas giant planets can form through direct gravitational collapse of the gas disc, if the disc is sufficiently dense. Here we show the formation prospects as a function of the disc's surface density at 1 au *(Σ)*, and viscosity parameter *(α)*. In the white region to the left, the disc is gravitationally stable out to 100 au. The colour scale and white contours show the minimum fragment mass in Jupiter masses (*30*), which would form at the inner edge of the unstable region. In the white region to the right, accretion rates are unphysically high (*33*). The red line marks the lower limit on *α* that gravitational instability itself would generate through turbulence. Hence, nearly all gravitationally unstable discs with realistic surface densities and accretion rates generate fragments in the mass range of the GJ 3512 planets. For reference, the minimum-mass solar nebula surface density value is around 1700 g cm⁻² (*34*).

# Supplementary Materials for

## A giant exoplanet orbiting a very-low-mass star challenges planet formation models

## Materials and Methods

### Spectroscopic observations and analysis

GJ 3512 was included in the list of targets of the CARMENES exoplanets survey (*13*) in late 2016. It was observed with the visual (5200-9600 Å) and near-infrared (9600-17100 Å) channels of the CARMENES spectrograph installed at the 3.5-m telescope at the Calar Alto Observatory (Almería, Spain) from December 2016 until May 2019, with an average cadence of 4 days, and 30-minute exposure times. The SERVAL (*19*) routine was used to obtain radial velocities for each échelle order and also several indices that can be used as stellar activity indicators. In this work we have made use of the radial velocities computed both from the visual (VIS) and the near-infrared (NIR) channels, for which a perspective (secular) acceleration of ~0.37 m s$^{-1}$ a$^{-1}$ is already corrected. Instrumental effects identified in CARMENES data, such as night-to-night offsets, are corrected using the large sample of stars observed (*35*), although the magnitude of such corrections is much smaller than the radial velocity variability of GJ 3512.

A total of 145 and 143 radial velocity epochs, provided in Tables S1 and S2, are available for GJ 3512 in the VIS and NIR channels, respectively. The median internal precision of the VIS data is about 2.5 m s$^{-1}$. In the NIR channel, several spectral orders (most of them in the ~1.2 μm band) contain very little radial velocity information (*18*). Radial velocities determined using those orders are therefore prone to systematics caused by telluric line contamination or unmasked detector defects. We carefully inspected the spectral orders to be used for radial velocity calculation. We tested various combinations of orders and selected the combination that minimized the radial velocity scatter of the full CARMENES M-dwarf sample. A total of 19 of the 55 available orders were used to achieve the best possible precision, obtaining a radial velocity curve in agreement with that of the VIS channel, with a median internal precision of about 7.1 m s$^{-1}$.

Radial velocity measurements were modelled using Keplerian orbits by maximization of the likelihood function (*36*). VIS and NIR data were fitted both individually and simultaneously. A jitter parameter was added in quadrature to the error bars of each dataset to account for an additional noise term as usually done in this context (*37*). We started with a single-planet Keplerian fit that reproduced reasonably well the large-amplitude radial velocity variability. However, a long-term trend was observable in the residuals. As a first approximation, we added a linear term to the fit, which improved the likelihood value by $\Delta \ln \mathcal{L} \sim 124$. We subsequently considered a second order polynomial for the fit that increased the significance even further ($\Delta \ln \mathcal{L} \sim 185$), indicating the presence of some curvature and therefore a tentative long-period signal. For this reason, we finally performed a recursive periodogram search (*38*) to constrain the properties of this additional signal in the data by fitting simultaneously a Keplerian function with a period of approximately 204 days and a sinusoidal function. Table S3 lists the properties of the different fits obtained. Fig. S1 displays the recursive periodogram as a function of the period of the long-term signal, and the derived properties of the outer object (i.e., radial velocity semi-amplitude, minimum mass and semi-major axis). The time span of the observations is not sufficient to determine this period, but the periodogram reveals that the data are consistent with a signal of period above ~1400 days. This makes this long-term trend consistent with a second object with a mass in the range between a Neptune-like planet and a brown dwarf, and semi-major axis between ~1 and ~10 au. High-resolution imaging taken during the definition of the CARMENES input catalogue (*39*) and also available in the bibliography (*40*) and the inspection of the high-resolution spectra to look for the possibility of the system being a spectroscopic binary exclude a stellar companion.

To check the consistency between the VIS and NIR radial velocities, individual fits were also obtained. The resulting parameters are listed in Table S4. The parameters of GJ 3512 b are consistent within the uncertainties in all cases. The long-period trend is less constrained in the case of the NIR radial velocity data due to larger uncertainties. However, the two-planet solution is still statistically more significant than a single Keplerian function, and the radial velocity semi-amplitude and the long-term trend are consistent for both wavelength channels. This favours the hypothesis of an additional planet in the system rather than the trend being caused by stellar activity. However, given the limited time baseline, we cannot fully exclude that part of the variation may be be caused by a long-period activity cycle, albeit there is no evidence in the various indicators.

Stellar activity indices are also retrieved from the CARMENES spectra and they are provided in Table S5. We searched for signals in the periodograms of the flux ratio of the Hα line, the differential line width and the chromatic index (*19*) of the VIS channel data. Fig. S2 shows the time-series and the likelihood periodograms of the radial velocities, the residuals of the best-fitting two-planet solution and the spectral indices. The periodogram of the radial velocity residuals shows peaks around ~92 and 340 days, although of low significance, which could be related to stellar activity and window aliases. We modelled the radial velocity data using Gaussian Process algorithms (*41*, *42*) to consider correlated noise. The resulting orbital parameters are consistent within the uncertainties with those reported in Tables 1 and S3. A 3-σ clipping procedure was applied to data from activity indices to remove outliers because some of these observations could be affected by flaring events. No significant signals above the significance level of $\Delta \ln \mathcal{L} > 15$ (~0.1% FAP) are found at the ~203-day period. Interestingly, the differential line width index shows a significant signal at a period of 88.8 days that is consistent with the rotation period derived from photometry.

Photometric monitoring

To better compare photometric and radial velocity variability signals due to stellar activity we obtained photometry contemporaneous with the CARMENES observations from different sites. The target was observed with the robotic 0.8-m Joan Oró telescope at the Montsec Astronomical Observatory (Lleida, Spain) using the standard Johnson-Cousins *R* band from December 2017 to May 2019. A total of 161 epochs were obtained using the MEIA2 instrument, a 2k×2k Andor CCD camera. Each epoch consisted in two to three blocks of observations with 15 s exposure time separated by few hours. These observations were combined to obtain a single nightly measurement to have the same cadence for all photometric data. This dataset was separated in two seasons to avoid any systematic caused by an intervention in the telescope in between.

We also collected photometric data in the Johnson $V$, $R$, and $I$ bands, and in the $V$ and $R$ bands with the T150 (1.5-m) and T90 (0.9-m) telescopes at Sierra Nevada Observatory (SNO, Granada, Spain), respectively. Both telescopes are equipped with similar CCD cameras VersArray 2k×2k (*43*). The observations with the T150 telescope were collected at 21 different epochs between February and May 2018. Each epoch typically consisted of 20 observations of 100 s, 50 s, and 20 s in the $V$, $R$, and $I$ bands, respectively, per night. The observations with the T90 telescope correspond to 33 epochs in the period October 2018 to January 2019. Each epoch usually consisted of 20 individual exposures of 150 s and 120 s in the $V$ and $R$ bands, respectively. A number of nearby and relatively bright stars within the frames were selected as check stars in order to choose the best ones to be used as reference stars. The same set of reference stars was chosen for both telescopes.

Finally, photometric observations of GJ 3512 were obtained in the $I'$ band with the 40-cm telescopes of Las Cumbres Observatory (LCO) at the Teide site. The observations yielded 16 different epochs in the interval between 5 and 28 March 2018. We typically acquired 60 individual exposures of 40 s in each epoch. Relative photometry of GJ 3512 was obtained with respect to several reference field stars of similar brightness.

Photometric measurements are provided in Table S6 and Fig. S3 shows the photometric time-series and their likelihood periodograms. There is a clear modulation of ~15 mmag and several signals are above the significance level of $\Delta \ln \mathcal{L} > 15$ (~0.1% FAP) for the different datasets. Assuming a sinusoidal modulation, the TJO first-season data are best fitted using a ~100-day period, while for the second season roughly significant signals at ~87 and ~42 days (after subtraction of the first signal) are present. On the other hand, the SNO data results in a period estimate of ~44 days for T90 observations, while a non-significant signal at ~73 days is present in T150 data. The different periods, which are close to the 2:1 ratio, may indicate an evolving spot pattern on the surface of the star, with active regions at opposite longitudes. LCO data do not show any clear signal due to the short time span of the observations. In order to combine all the datasets corresponding to the same photometric band, we first removed a mean value from each light curve and we then computed zero point offsets for each SNO T150 and T90 $R$ band datasets with respect to TJO first and second seasons, respectively, and LCO $I'$ band dataset with respect to SNO T150 $I$ band data, by averaging the observations corresponding to the same night. The combined datasets yield significant periods at ~44 and ~84 days in the $V$ band, ~87 and ~42 days in $R$ band, and ~74 days in the $I$ band data. Fitting a zero point between the two seasons of observations for $V$ and $R$ band datasets produces similar results. Therefore, we conclude that the rotation period of GJ 3512 is probably around ~87 days, which agrees very well with the signal found in spectroscopic activity indicators such as the differential line width. From the combined

*R* band light curve, we can estimate a rotation period of 87 ± 5 days, with the uncertainty estimated from the width of the periodogram peak.

GJ 3512 was also observed by the MEarth (*44*) and SuperWASP (*45*) surveys. MEarth observations span between October 2008 and June 2017, including 3400 epochs and a dispersion of ~8 mmag. The likelihood periodogram of the datasets shows a peak at ~80 days but with a semi-amplitude modulation of only 3 mmag, much smaller than those obtained from other observations and consistent with the value reported in ref. *20*. Interestingly, the two most densely sampled seasons, between October 2008 and June 2009 (131 nights) and September 2012 and June 2013 (165 nights), show variability with a period of ~41 and ~75 days and semi-amplitudes of ~5.5 and ~3 mmag, respectively, in agreement with the values reported above. This indicates that the spot pattern on the photosphere of the star varies over time and that the modulation caused by spots is currently more apparent than 5-10 years ago. On the other hand, SuperWASP data only comprise 16 nights of observations between March 2007 and April 2008 with a dispersion of about 0.1 mag, which is not useful for our analysis.

Age estimate

The non-detection of Li I in the spectra indicates that GJ 3512 is not a very young star. Furthermore, its long rotation period points at a rather old age for the system. The observed space motions of the star are consistent with GJ 3512 being part of the thin disc population of the Galaxy, for which ref. *46* estimates an age in the range 6.8 – 8.2 Ga based on the analysis of nearby white dwarfs. Following the method applied to the TRAPPIST-1 planetary system (*47*), we estimate an age in the interval 3 – 9 Ga from the *UVW* velocities. These are calculated from the Gaia proper motions and distance, and from the radial velocity measured using CARMENES spectra of +6.2 ± 0.5 km s$^{-1}$. Current gyrochronology relationships do not extend to stars as cool as GJ 3512. If we extrapolate the calibration by (*48*) to the measured colour index *B – V* = 1.84 (*49*) mag of GJ 3512, we obtain an age in the interval 4 – 7 Ga. We caution, however, that this determination can be biased, although the agreement with the method using space motions is good. From our analysis, we constrain the age to be likely within 3 – 8 Ga.

Formation models of the planetary system

**Planet formation by pebble accretion.** We first consider planet formation in the pebble accretion paradigm (*50*). Here, planetary cores are built up by the efficient accretion of cm-sized particles, aided by gas drag; following which, if the core becomes large enough, gas accretion leads to growth to Jovian sizes. This formation model has been shown to reproduce well the diversity of planets seen around Solar-type stars, with the final mass and orbital radius depending on parameters such as the formation site and time of the planetary cores, and the pebble flux through the disc (*51*). Our formation model is adapted from the latest published pebble accretion models (*24*), and considering several host star parameters, in particular, a stellar mass (0.1 M$_\odot$) and luminosity (0.1 L$_\odot$) approximately matching GJ 3512's pre-main sequence properties (*52*). Planetary seeds of 0.01 M$_\oplus$ are inserted into the protoplanetary disc at time $t_0$ and radius $r_0$, and experience accretion of pebbles and gas as well as migration torques, until the disc dissipates at 3 Ma (*53*). The seeds are assumed to have formed directly as large planetesimals from the streaming instability or by an earlier phase of growth by planetesimal accretion and pebble accretion (*54*). In Fig. S5 we show the final position and mass of planets formed as a function of $t_0$ and $r_0$ for four stellar masses including the adopted values for GJ 3512, with the disc accretion rate $\propto M^2$, starting at an accretion rate of $10^{-7}$ $M_\odot$ $a^{-1}$ $(M_\star/M_\odot)^2$ at t = 0 and dissipating at an accretion rate of $10^{-8}$ $M_\odot$ $a^{-1}$ $(M_\star/M_\odot)^2$ three million years later. As expected, a wide range of planets is successfully formed around the more massive stars. However, planetary seeds around the 0.1-$M_\odot$ star experience only modest growth, due to a higher scale-height of the pebble layer (lower stellar mass) and a lower pebble column density in these low-mass discs. At the same time, migration is faster when the central star is lighter. The result is that cores either remain light due to the low pebble accretion rates, if they form far from the star, or migrate rapidly towards the inner disc edge before growing appreciably by pebble accretion, if they start further in where the growth rate is higher. Besides this model (hereafter model 1), we further explored models where the initial gas accretion rate is $3\times10^{-7}$ $M_\odot$ $a^{-1}$ and the final gas accretion rate is either $10^{-10}$ $M_\odot$ $a^{-1}$ (model 2) or $2\times10^{-9}$ $M_\odot$ $a^{-1}$ (model 3) for a stellar host star of 0.1 $M_\odot$. The initial total gas masses in these three models are $M_{\mathrm{disc}}$ = 0.0005 $M_\odot$, 0.009 $M_\odot$, and 0.06 $M_\odot$, respectively. The results are shown in Fig. S6. Here we use a logarithmic time axis to emphasize the earliest stages of the protoplanetary disc evolution. The yellow line shows the size of the protoplanetary disc (which expands in time due to outwards transport of angular momentum), the red contours show the Toomre Q parameter for gravitational instability in the gas, while the white contours and the background colour show the final position and mass of the planet, respectively. Model 3 is gravitationally unstable between approximately 20 and 30 au (the approximate outer edge of the disc), but we do not include planet formation by disc instability in these models. All the models fail to produce gas giant planets by pebble accretion and gas accretion.

**Planetesimal accretion.** In the previous section on planet formation by pebble accretion we assumed that planetesimal accretion did not contribute to the growth of the cores. Here we include also planetesimal accretion (*55*). We assume that 1% of the initial gas column density profile is converted to planetesimals of 100 km in radius early on. The planetesimal accretion rate is calculated from the model in ref. *56* describing how a migrating protoplanet accretes a fraction of the planetesimals that it plows through. In Fig. S7 we show again models 1, 2, 3, as described in the previous section, but now including planetesimals. Clearly, planetesimals do not contribute noticeably to the nominal model where the gas column density is low and hence the initial planetesimal population of relatively low mass. However, for the two dense protoplanetary disc models (model 2 and model 3), planetesimal accretion leads to the formation of gas-giant planets that end up migrating very close to the host star (in contrast to the companions of GJ 3512). The reason why planetesimal accretion is important lies in our assumption that the young, compact protoplanetary disc converts about 1% of its local mass to planetesimals. Model 3 initially forms 150 $M_\oplus$ of planetesimals out to 20 au; scaling this to a solar-mass star corresponds to the conversion of 1500 $M_\oplus$ of dust to planetesimals. This is a debatable assumption for GJ 3512. The central star has solar metallicity and that was likely the case for the protoplanetary disc as well. Triggering planetesimal formation by the streaming instability requires a minimum of 1.5–2% mass loading of pebbles in the gas (*57*). Such metallicity levels could occur near ice lines (*58*) and by late photoevaporation of gas (*59*). It is possible that stars of very high metallicity could experience an early transformation of the dense, circumstellar gas to a population of planetesimals, but such a scenario is likely not relevant for the metallicity of GJ 3512. On the other hand ref. *60* showed that planetesimal formation in trapping mode, when the local pebble flux controls the planetesimal formation in traps like zonal flows and vortices, has no such severe metallicity constraints and furthermore forms planetesimals in the inner nebula from pebbles at larger distances, thus the local planetesimal to gas ratio can be much larger than the original dust to gas ratio. Yet, the mentioned models were performed for solar type stars and it still has to show how this prediction will translate to M-dwarfs. We therefore conclude that current core accretion models (either pure pebble accretion or combined pebble and planetesimal accretion) have severe difficulties to explain GJ 3512.

**Planet formation by gravitational instability.** Given the conclusions in the previous section, we turn to the alternative scenarios and consider that the planet orbiting GJ 3512 could have formed by spontaneous gravitational fragmentation of the gas disk (*31*). We work with an analytical disc model (*61*) and include active heating from the disc's viscosity and passive heating from the luminosity of the central star, additionally imposing a minimum temperature of 10 K to account for heating from other stars in the birth cluster. We construct a grid of such disc models, varying the viscosity parameter $\alpha$ and the mass accretion rate $dM/dt$. We take as an upper limit on the accretion rate $10^{-4}$ $M_\odot$ $a^{-1}$, the upper limit measured for embedded Class I young stellar objects (*62*) and FU Orionis objects (*63*). Numerical simulations show that a gravitationally unstable disc generates an $\alpha$ of $0.01 - 1$ (*64*). For each disc, over radii from 0.1 to 100 au, we calculate Toomre's stability parameter $Q = 3\alpha c_s^3/(G \, dM/dt)$, where $c_s$ is the sound speed and $G$ the gravitational constant (*31*). $Q$ decreases outwards in these disc models, and when $Q \lesssim 1$ the discs become gravitationally unstable and fragment. The mass that should be taken by these fragments, which undergo gravitational collapse to form giant planets directly, is disputed in the literature, and we use the analytical scaling of ref. *30*, which agrees with some numerical simulations (*65*). These are the masses shown in Fig. 3, and they encompass the observed minimum mass of GJ 3512 b.

**Planet formation by prompt fragmentation of the molecular cloud.** Since GJ 3512 b stretches the limits of the core accretion as well as the disc fragmentation scenarios, one could consider a third formation mechanism, namely prompt fragmentation of a filament in the natal molecular cloud. This would be analogous to the formation of binary and triple stars, but again with extreme parameters, including a mass ratio $q < 0.01$. Simulations of star formation in clusters do not produce such systems (*66*); therefore, this formation channel cannot be considered likely.

**System formation by capture.** Given the challenges of forming such massive planets around this star, we also considered the scenario where the planets were captured by GJ 3512 during a close encounter with another star in its birth cluster (*67*). However, this scenario can be ruled out for several reasons. First, capture of planet by a star requires a flyby pericentre $\lesssim 3$ times the planet's semi-major axis, or around 1 au for GJ 3512 b. Flybys this close are rare, with a rate of $\sim 10^{-4}$ M $a^{-1}$ in a young cluster (*68*). Second, capture rates by low-mass stars are lower (*67*), and even capturing two planets is considerably less likely than capturing only one. Finally, the observed eccentricity of $\sim 0.43$ is fairly low for captured planets (*67*).

## Early dynamical evolution of the planetary system

The orbital eccentricities of planets embedded in a disc are usually damped. Planet–planet scattering has been shown to reproduce well the eccentricity distribution of observed giant exoplanets (*7*), therefore we present here some possible dynamical histories of the GJ 3512 system. Because of the possible presence of a second, outer planet, and the fact that an unstable two-planet system will end up losing a planet through collision or ejection, we consider systems of initially three planets. A moderately large eccentricity of 0.43 cannot be achieved by the ejection of too small a planet (*69*), and so we construct systems of planets in a 2:1:4 mass ratio. Other configurations are possible and will be explored in a follow-up study. We set up 1000 systems where the planets are separated by 3 to 5 mutual Hill radii, close to the 4:2:1 mean motion resonance, and integrate their orbits with the RADAU integrator in the MERCURY (*70*) package for $10^8$ years. After this time, 733 of the systems have experienced instability and lost at least one planet, 524 possessing two planets at the end of the simulation. Of these, 124 have one planet with orbital elements similar to those of GJ 3512 b ($a < 0.6$ au; $0.33 \leq e \leq 0.53$). The eccentricities and semi-major axes of planets in these systems are shown in Fig. S8. The outer planets in these systems typically lie on lower eccentricity orbits between 1 and 2 au and are consistent with the long-term trend observed in the data.

**Table S1. Radial velocity time-series from the CARMENES visual channel spectra.** Only a subset of the data analysed in this paper is shown here. A machine-readable version of the full dataset is available in Data S1.

| Barycentric Julian Date, BJD | Radial velocity, RV (m s$^{-1}$) |
|:---:|:---:|
| 2457752.40306 | −89.8±1.7 |
| 2457755.38261 | −92.6±2.8 |
| 2457761.47313 | −94.1±2.6 |
| 2457788.49993 | −43.1±3.6 |
| 2457815.46609 | −14.1±2.0 |
| 2457823.56615 | −7.2±1.9 |
| 2457833.40010 | 0.3±2.9 |
| 2457848.36227 | 16.3±2.0 |
| 2457852.39121 | 20.2±3.0 |
| 2457856.41314 | 21.7±2.3 |

**Table S2. Radial velocity time-series from the CARMENES near infrared channel spectra.**
Only a subset of the data analysed in this paper is shown here. A machine-readable version of the full dataset is available in Data S1.

| Barycentric Julian Date, BJD | Radial velocity, RV (m s$^{-1}$) |
|---|---|
| 2457752.40244 | −84.2±6.1 |
| 2457755.38269 | −97.0±6.2 |
| 2457761.47294 | −97.1±6.0 |
| 2457788.49882 | −55.2±9.1 |
| 2457815.46603 | −11.3±6.2 |
| 2457823.56598 | −16.8±6.7 |
| 2457833.40043 | 0.1±7.8 |
| 2457848.36220 | 9.8±6.2 |
| 2457852.39119 | 38.6±9.5 |
| 2457856.41301 | 10.6±5.2 |

**Table S3. Parameters of the radial velocity simultaneous fits to VIS and NIR channels.** The likelihood ($\mathcal{L}$) value of the single Keplerian fit is used as reference (ln $\mathcal{L} = -1067.51$). A long-period trend is present in the data, but the parameters of the tentative planet GJ 3512 c cannot be further constrained, thus only lower limits of the period and radial velocity semi-amplitude, and the range of allowed phase differences ($\theta$) with respect to GJ 3512 b periastron passage are given.

| | | Model | | |
|---|---|---|---|---|
| **Fit parameters** | | **Keplerian** | **Keplerian+Linear** | **Keplerian+circular** |
| $P_b$ (d) | | $204.23 \pm 0.32$ | $204.04 \pm 0.21$ | $203.59 \pm 0.14$ |
| $K_b$ (m s$^{-1}$) | | $69.02^{+0.84}_{-0.87}$ | $71.14 \pm 0.48$ | $71.84^{+0.34}_{-0.36}$ |
| $e_b$ | | $0.444 \pm 0.010$ | $0.4269^{+0.0059}_{-0.0061}$ | $0.4356 \pm 0.0042$ |
| $\omega_b$ (deg) | | $125.6^{+1.8}_{-1.7}$ | $125.0 \pm 1.0$ | $125.49^{+0.71}_{-0.73}$ |
| $T_{\text{periastron,b}}$ (BJD-2450000.0) | | $7744.4 \pm 1.2$ | $7744.34^{+0.80}_{-0.78}$ | $7745.65 \pm 0.50$ |
| $P_c$ (d) | | ... | ... | $>1390$ |
| $K_c$ (m s$^{-1}$) | | ... | ... | $> 12$ |
| $\theta$ (deg) | | ... | ... | $60 - 90$ |
| $\gamma$ (m s$^{-1}$) | VIS | $-0.76^{+0.67}_{-0.66}$ | $4.6^{+3.2}_{-1.2}$ | $4.6^{+3.2}_{-1.2}$ |
| | NIR | $0.0 \pm 1.0$ | $-6.5^{+3.2}_{-1.6}$ | $-6.5^{+3.2}_{-1.6}$ |
| $S$ (m s$^{-1}$ a$^{-1}$) | | ... | $12.02^{+0.58}_{-0.56}$ | ... |
| jitter (m s$^{-1}$) | VIS | $7.51^{+0.53}_{-0.51}$ | $4.7^{+1.3}_{-1.4}$ | $1.38^{+0.36}_{-0.41}$ |
| | NIR | $9.17^{+0.98}_{-0.88}$ | $5.128^{+0.97}_{-0.89}$ | $4.7^{+1.3}_{-1.4}$ |
| **Fit quality** | | **Value** | | |
| ln $\mathcal{L}_{\text{model}}$ − ln $\mathcal{L}_{\text{keplerian}}$ | | $0$ | $124.09$ | $187.57$ |
| rms (m s$^{-1}$) | VIS | $7.88$ | $4.14$ | $3.27$ |
| | NIR | $12.71$ | $10.99$ | $10.26$ |

**Table S4. Parameters of the radial velocity fits to the separate and combined VIS and NIR data sets.** The likelihood value of the single Keplerian fit is used as reference: ln $\mathcal{L}$ = −505.21 (VIS), ln $\mathcal{L}$ = −560.43 (NIR) and ln $\mathcal{L}$ = −1067.51 (VIS and NIR simultaneously). The long-period trend is detected in the VIS channel radial velocity data, while the NIR data fit likelihood is only marginally better than for a single Keplerian fit. Only lower limits of the period and radial velocity semi-amplitude, and the range of allowed phase differences ($\theta$) with respect GJ 3512 b periastron passage are given.

| | | Data set | | |
|---|---|---|---|---|
| **Fit parameters** | | **VIS** | **NIR** | **VIS+NIR** |
| $P_b$ (d) | | $203.61\pm0.14$ | $203.29\pm0.37$ | $203.59\pm0.14$ |
| $K_b$ (m s$^{-1}$) | | $71.50^{+0.39}_{-0.36}$ | $74.1\pm1.1$ | $71.84^{+0.34}_{-0.36}$ |
| $e_b$ | | $0.4337\pm0.0044$ | $0.451\pm0.013$ | $0.4356\pm0.0042$ |
| $\omega_b$ (deg) | | $125.76^{+0.76}_{-0.75}$ | $123.9^{+1.9}_{-2.0}$ | $125.49^{+0.71}_{-0.73}$ |
| $T_{\text{periastron,b}}$ BJD2450000.0 (d) | | $7745.61^{+0.53}_{-0.50}$ | $7746.2\pm1.3$ | $7745.65\pm0.50$ |
| $P_c$ (d) | | > 1410 | >920 | >1390 |
| $K_c$ (m s$^{-1}$) | | > 12 | > 13 | > 12 |
| $\theta$ (deg) | | 60 – 90 | 45 – 90 | 60 – 90 |
| $\gamma$ (m s$^{-1}$) | VIS | $17.18^{+17.1}_{-4.0}$ | $10.9^{+3.7}_{-1.5}$ | $4.6^{+3.2}_{-1.2}$ |
| | NIR | ... | $10.70^{+3.7}_{-1.8}$ | $-6.5^{+3.2}_{-1.6}$ |
| jitter (m s$^{-1}$) | VIS | $1.21^{+0.37}_{-0.44}$ | ... | $1.38^{+0.36}_{-0.41}$ |
| | NIR | ... | $1.23^{+0.38}_{-0.45}$ | $4.7^{+1.3}_{-1.4}$ |
| **Fit quality** | | Value | | |
| ln $\mathcal{L}_{\text{model}}$ − ln $\mathcal{L}_{\text{keplerian}}$ | | 143.31 | 47.74 | 187.57 |
| rms (m s$^{-1}$) | VIS | 3.25 | ... | 3.27 |
| | NIR | ... | 10.03 | 10.26 |

**Table S5. Time-series of the stellar activity indices derived from the CARMENES visual channel spectra.** All indices are computed using SERVAL (*20*). Only a subset of the data analysed in this paper is shown here. A machine-readable version of the full dataset is available in Data S1.

| Barycentric Julian Date, BJD | Hydrogen Balmer line ratio, H $\alpha$ | Differential line width, dLW (m$^2$ s$^{-2}$) | Chromatic index, CRX (m s$^{-1}$) |
|---|---|---|---|
| 2457752.40306 | 1.747$\pm$0.009 | $-3.9\pm$2.3 | 28.1$\pm$14.5 |
| 2457755.38261 | 1.653$\pm$0.012 | $-5.8\pm$2.7 | $-70.5\pm$28.1 |
| 2457761.47313 | 1.325$\pm$0.009 | $-11.3\pm$2.2 | 47.4$\pm$26.6 |
| 2457788.49993 | 1.503$\pm$0.018 | $-22.8\pm$5.2 | $-5.5\pm$38.8 |
| 2457815.46600 | 2.391$\pm$0.011 | $-21.1\pm$2.8 | 34.5$\pm$17.1 |
| 2457823.56615 | 1.875$\pm$0.008 | $-6.5\pm$1.9 | 36.3$\pm$14.1 |
| 2457833.40010 | 1.805$\pm$0.014 | $-15.6\pm$3.2 | 33.7$\pm$29.7 |
| 2457848.35684 | 2.091$\pm$0.009 | $-5.0\pm$2.5 | 27.1$\pm$15.8 |
| 2457852.39121 | 1.709$\pm$0.010 | 0.0$\pm$3.5 | 55.4$\pm$28.5 |
| 2457856.41314 | 1.677$\pm$0.008 | $-1.4\pm$2.4 | 19.9$\pm$21.6 |

**Table S6. Photometric light curves of GJ 3512 obtained at the Montsec, Sierra Nevada and Las Cumbres observatories.** Only a subset of the data analysed in this paper is shown here. A machine-readable version of the full dataset is available in Data S1.

| Barycentric Julian Date BJD | Differential photometry (mag) | Photometric band | Telescope/instrument |
|---|---|---|---|
| 2458091.48957 | −0.0482±0.0075 | R | TJO/MEIA2 |
| 2458092.58100 | −0.0468±0.0050 | R | TJO/MEIA2 |
| 2458093.57870 | −0.0491±0.0041 | R | TJO/MEIA2 |
| 2458094.65474 | −0.0437±0.0050 | R | TJO/MEIA2 |
| 2458095.52833 | −0.0435±0.0059 | R | TJO/MEIA2 |
| 2458096.57270 | −0.0472±0.0030 | R | TJO/MEIA2 |
| 2458103.58412 | −0.0477±0.0039 | R | TJO/MEIA2 |
| 2458104.52712 | −0.0443±0.0041 | R | TJO/MEIA2 |
| 2458106.50138 | −0.0486±0.0048 | R | TJO/MEIA2 |
| 2458107.48880 | −0.0386±0.0035 | R | TJO/MEIA2 |

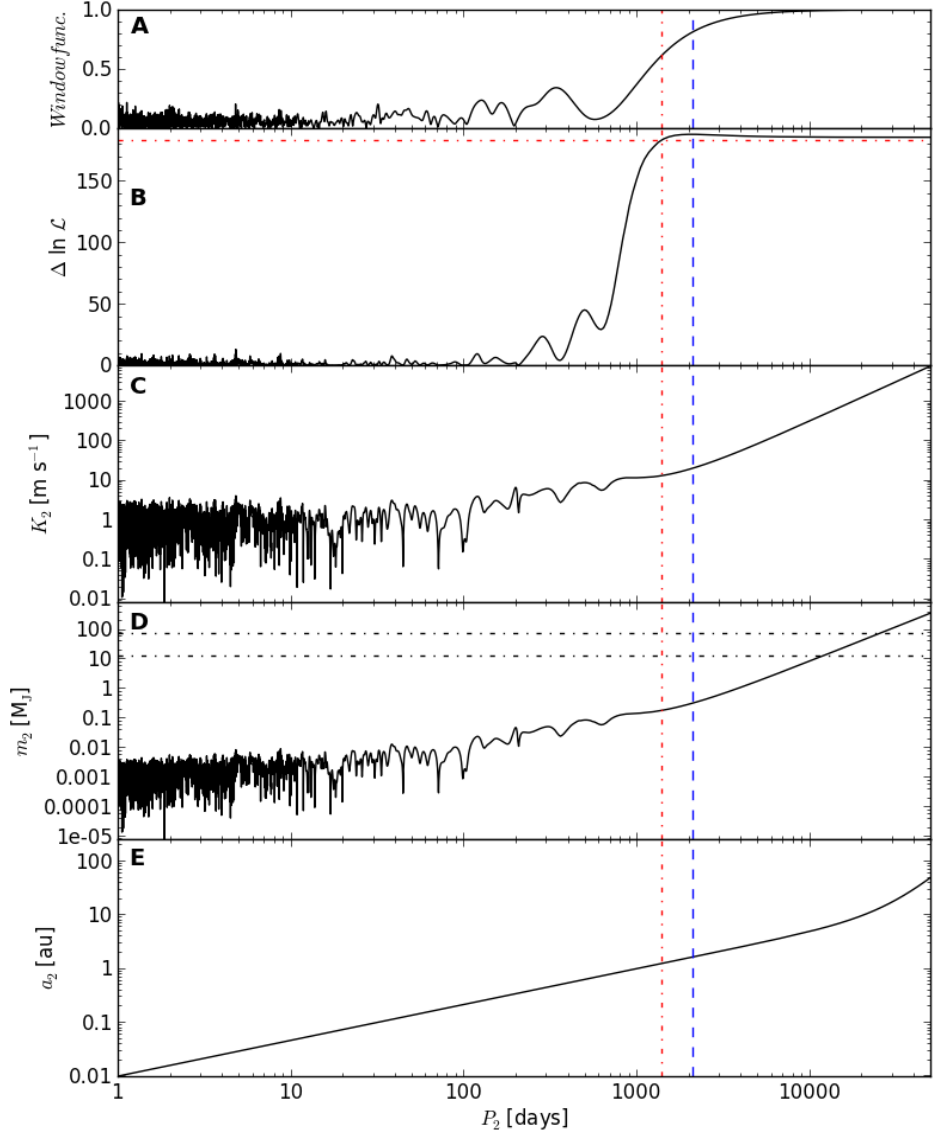

**Fig. S1. Recursive periodogram to search for additional signals in the time series of radial velocity data and residuals.** Panel A shows the window function of the radial velocity time series obtained with CARMENES. The $\ln \mathcal{L}$ difference of the two planets fits with respect to a single Keplerian fit as a function of the period of the long-term trend ($P_2$) is displayed in panel B. Panels C, D and E show the radial velocity semi-amplitude, the estimated mass, and the orbital semi-major axis corresponding to each long-period fitted, respectively. The dashed blue vertical line marks the position of the best likelihood solution (corresponding to a period of ~2100 days), and the dot-dashed red line marks the lower threshold of acceptable solutions according to their likelihood values ($\ln \mathcal{L}_{max} - \ln \mathcal{L}_{threshold} = 5$).

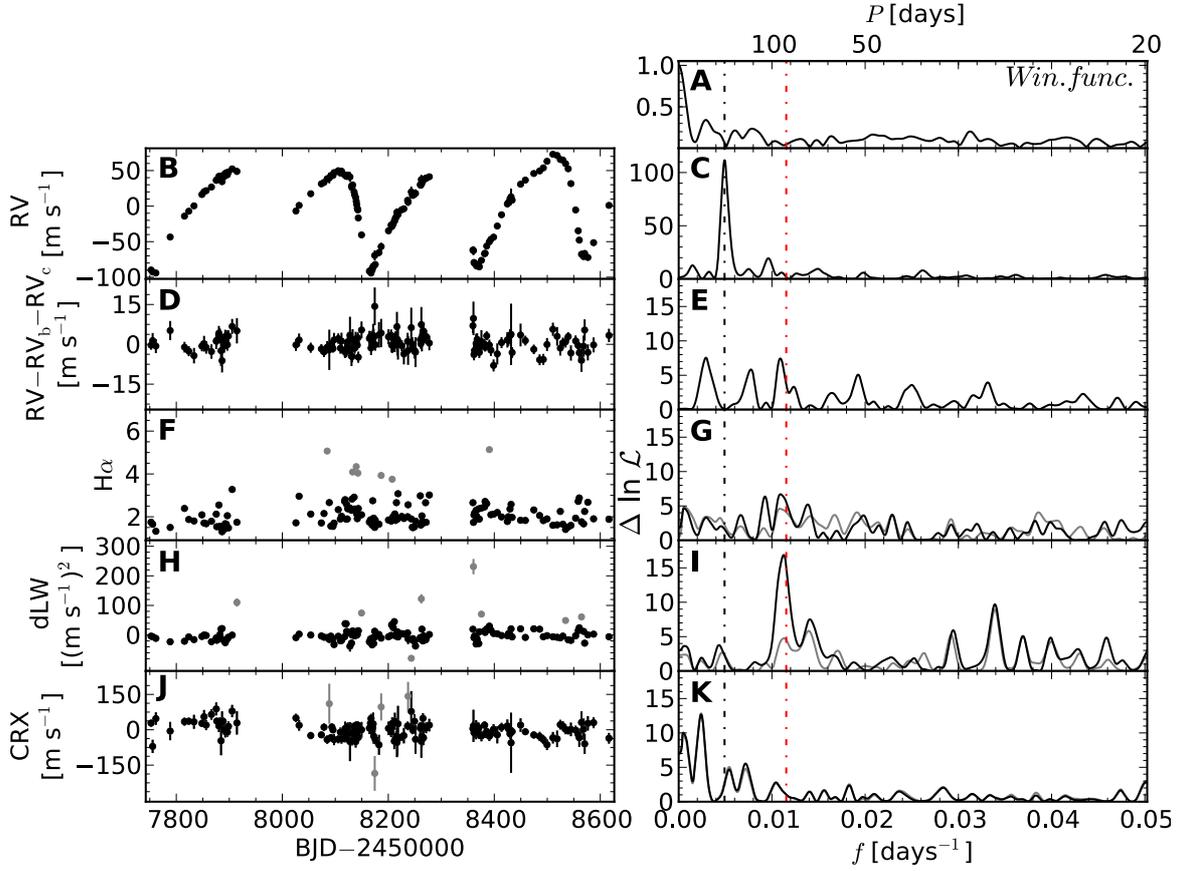

**Fig. S2. Radial velocity and activity indices obtained from the VIS channel CARMENES observations.** The left panels show the timeline of the measurements, with grey symbols indicating outliers removed by applying a 3σ clipping. Some of these may be caused by flaring events. The right panels show the likelihood periodograms of the full dataset (grey line) and with outliers removed (black line). The window function of the observations is shown in the top right panel. False alarm probability of 0.1% typically corresponds to $\Delta \ln \mathcal{L} \sim 15$. Therefore, no significant signals are found in stellar activity indices near the orbital period corresponding to GJ 3512 b (dot-dashed black vertical line). Significant signals with false alarm probabilities below or close to 1% are only present in the case of the differential line width (dLW) and chromatic index (CRX), showing peaks at ~89 days, that can be attributed to the rotation period of the star (dot-dashed red vertical line, ~87 days), and at ~415 days, that could be due to the yearly alias of a decreasing trend, respectively.

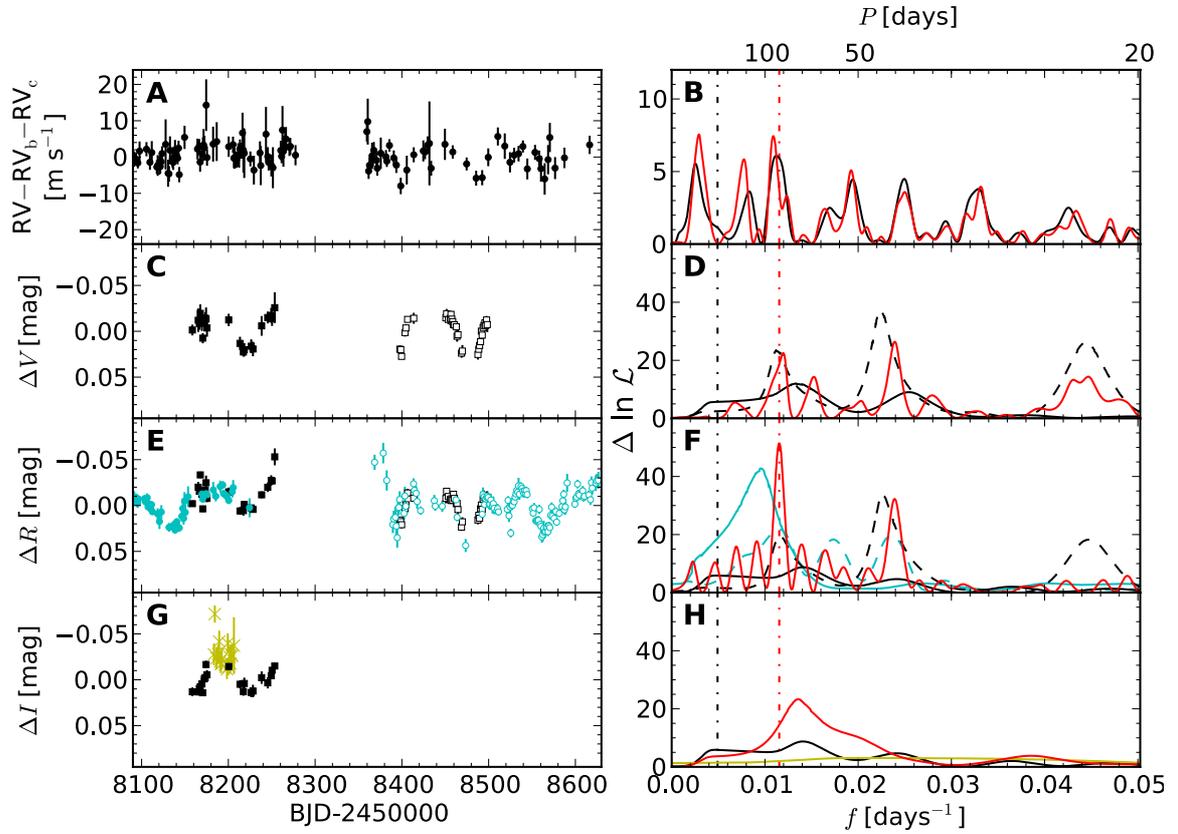

**Fig. S3. CARMENES VIS channel residual radial velocity and contemporaneous photometric time-series of GJ 3512.** The left panels display the radial velocity after subtraction of the planetary and long-term signals (black dots, panel A), along with the TJO (cyan solid and open dots), SNO T150 (black solid squares), SNO T90 (black open squares), and LCO (yellow crosses) photometric data. Panels C, E, and G correspond to different photometric bands as labelled. Zero points between TJO and SNO and SNO T150 and LCO light curves were computed from the mean value of measurements taken during the same night. The likelihood periodogram of each dataset is plotted in the right panels using the same colours (solid and dashed lines correspond to solid and open symbols, respectively), while the red line corresponds to the periodogram of the whole sample of radial velocity residuals (panel B) or of the combined photometric dataset (panels D, F, and G). False alarm probability of 0.1% typically corresponds to $\Delta \ln \mathcal{L} \sim 15$. A different vertical scale is used in panel B. Black and red dot-dashed vertical lines indicate the orbital period corresponding to GJ 3512 b and the estimated rotation period (~87 days), respectively.

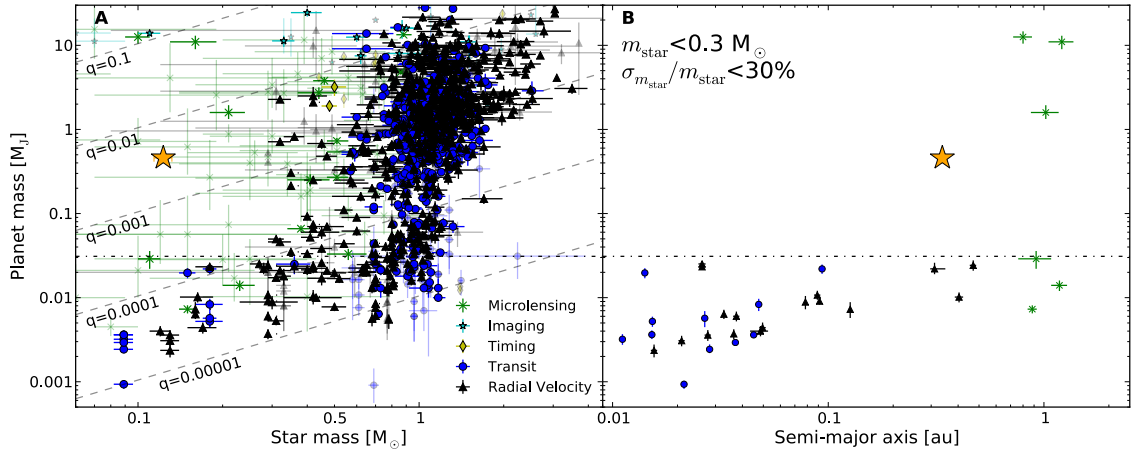

**Fig. S4. GJ 3512 b minimum planet mass and host stellar mass compared with known planetary systems.** Data for known exoplanets comes from the NASA exoplanet archive. Different exoplanet detection techniques are shown as labelled and GJ 3512 b is depicted with an orange star symbol. Panel A shows exoplanet masses as a function of the stellar host star masses. The planet minimum mass is plotted in the case of planets detected by radial velocities and timing. Dashed lines indicate different host star-to-planet mass ratios ($q$) as labelled, and the horizontal dot-dashed line corresponds to 10 M$_\oplus$. Systems with stellar or planet mass uncertainties above 30% (not shown in Fig. 2) are displayed in lighter colours. Panel B displays the exoplanet mass versus the orbital semi-major axis for late-type stars with masses below 0.3 M$_\oplus$ and relative uncertainties below 30%.

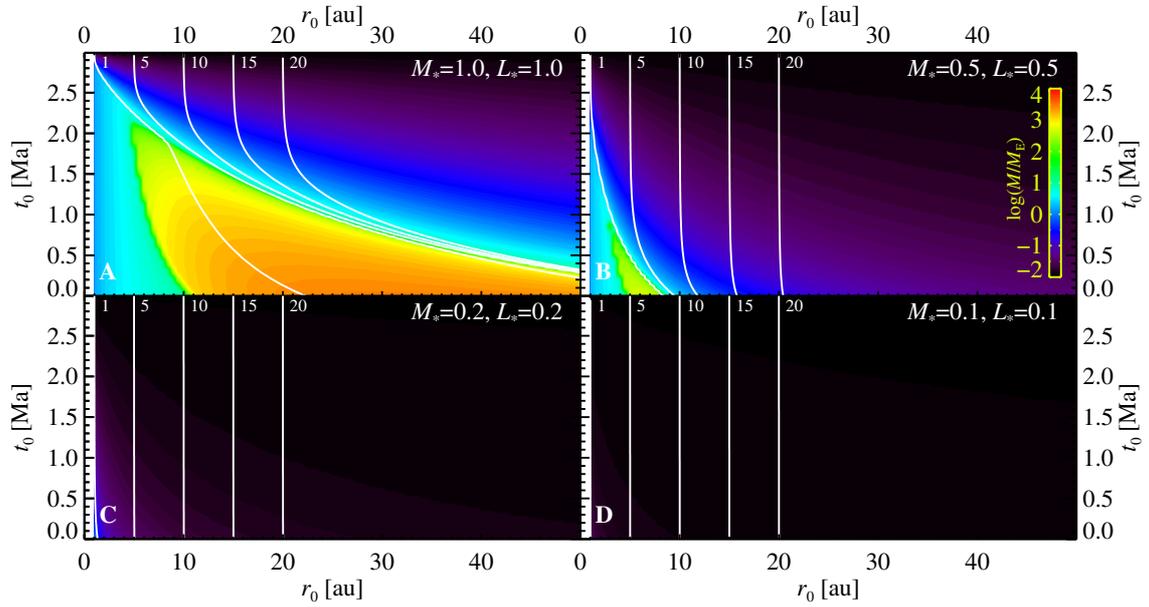

**Fig. S5. Formation maps for the pebble accretion scenario.** Each panel shows planet formation outcomes for a different stellar mass. In each panel, the x-axis shows the formation location of the 0.01 $M_\oplus$ seed, and the y-axis its formation time. The colour scale shows the final mass of the planet, and the black contours its final orbital radius (in au). The formation of a diverse spectrum of planets is possible around the more massive stars, but the seeds placed around 0.1- $M_\odot$ M-dwarfs do not significantly accrete or migrate. Stellar parameters are in solar units.

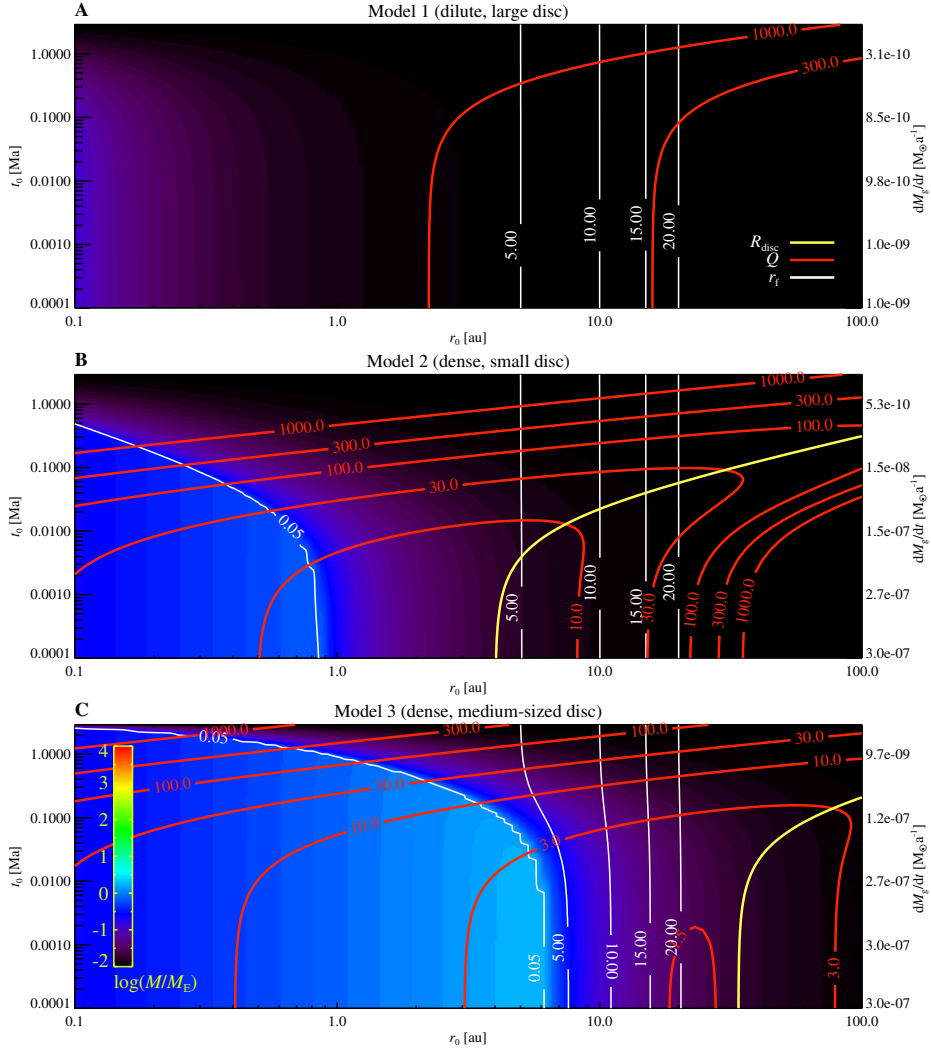

**Fig. S6. Growth maps for three protoplanetary disc models around a star of mass 0.1 M$_\oplus$.**
Model 1 (panel A) has an initial gas accretion rate of $10^{-9}$ M$_\oplus$ a$^{-1}$ and a final gas accretion rate of $10^{-10}$ M$_\oplus$ a$^{-1}$. Model 2 (panel B) has an initial gas accretion rate of $3\times10^{-7}$ M$_\oplus$ a$^{-1}$ and the same final gas accretion rate; this rapid decline of the accretion rate makes for an initially dense and compact disc. Finally, model 3 (panel C) has an initial gas accretion rate of $3\times10^{-7}$ M$_\oplus$ a$^{-1}$ and a final gas accretion rate of $2\times10^{-9}$ M$_\oplus$ a$^{-1}$, yielding a larger initial disc size compared to model 2. The yellow line shows the size of the protoplanetary disc (which expands in time due to outwards transport of angular momentum), the red lines mark the Toomre Q parameter for gravitational instability in the gas. Model 3 is gravitationally unstable in the gas between approximately 20 and 30 au (the initial disc size). The white contours indicate the final planetary position and the coloured contours the final planetary mass. Cores never grow above a few Earth masses, even in the gravitationally unstable model 3. This is mainly due to the rapid planetary migration around low-mass stars.

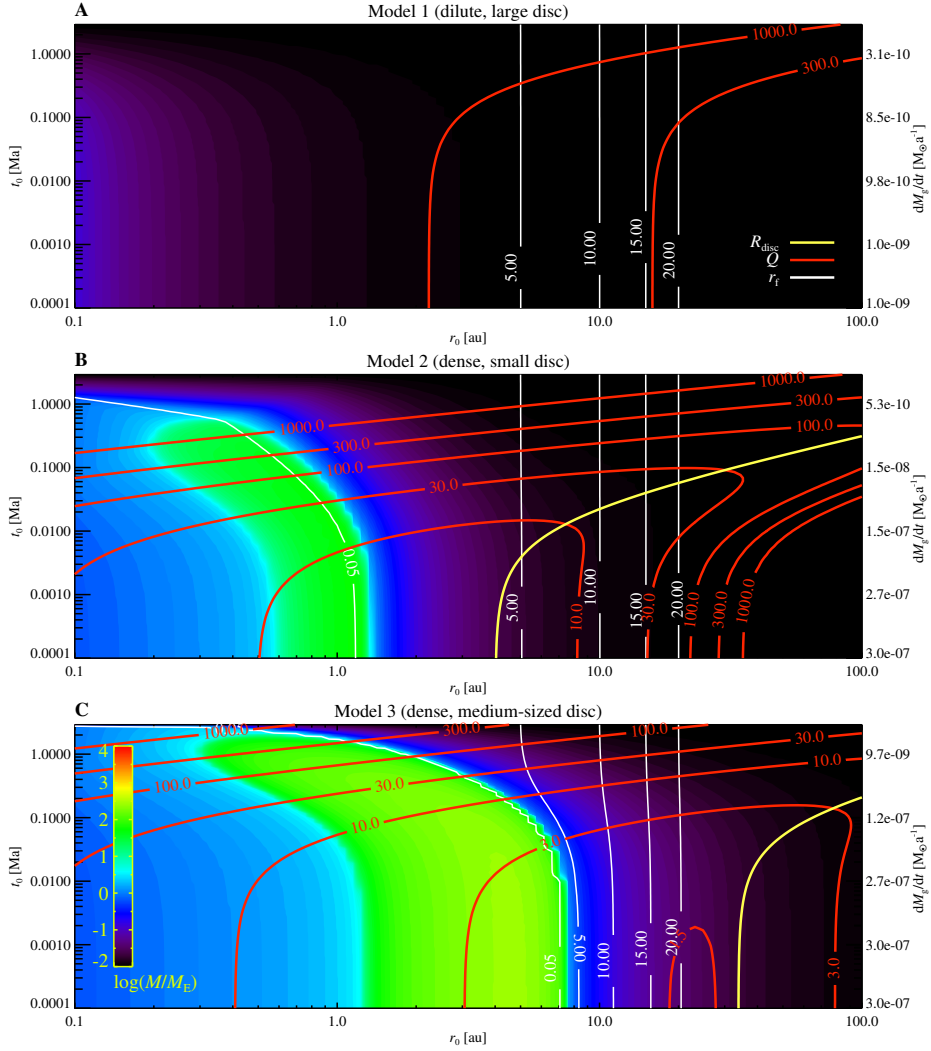

**Fig. S7. Growth maps for three protoplanetary disc models around a star of mass 0.1 M$_\oplus$, including planetesimal accretion.** We make the extreme assumption that the protoplanetary disc converts 1% of its initial mass at $t = 0$ to planetesimals, knowingly that this is not a realistic assumption for the solar-metallicity star GJ 3512. Planetesimal accretion by the migrating protoplanet leads to the formation of gas-giant planets in both models 2 and 3 (panels B and C, respectively). However, these giant planets migrate to orbits very close to the host star, becoming hot Jupiter-type planets, in contrast to the companions of GJ 3512.

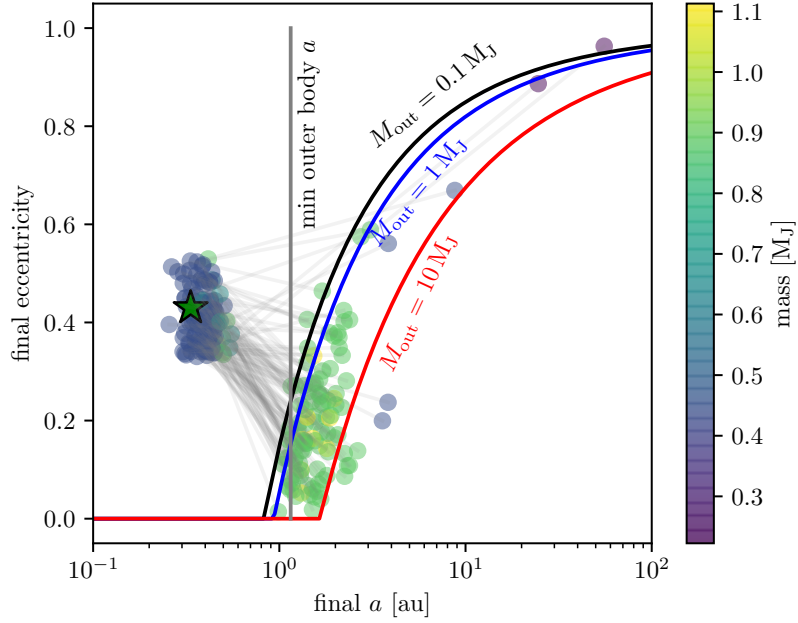

**Fig. S8. Semi-major axes *a* and eccentricities of planets in systems after scattering.** We show the orbital elements of the 124 out of 1000 systems in our dynamical simulation with two surviving planets and with the innermost planet having *a* < 0.6 au and *e* within 0.1 of the observed value. The vertical grey line marks the smallest possible orbit for the outer planet according to the radial velocity solution. Black, blue and red lines mark analytical estimates for the maximum permissible eccentricity for the outer planet of specified mass guaranteeing the long-term stability of two-planet systems (*71*). The green star symbol marks the orbital properties of GJ 3512 b (see Table 1). After ejecting one planet, most of the simulations yield an outer planet in a relatively low-eccentricity orbit (*e* = 0.1–0.3), a semi-major axis within 1–2 au and a mass of about 1 $M_J$. These parameters are compatible with the trend visible in the radial velocity observations.